# Networking and processing in optical wireless

**Osama Zwaid Alsulami[1], Amal A. Alahmadi[1], Sarah O. M. Saeed[1], Sanaa Hamid Mohamed[1], T. E. H. El-Gorashi[1], Mohammed T. Alresheedi[2] and Jaafar M. H. Elmirghani[1]**

[1]*School of Electronic and Electrical Engineering, University of Leeds, LS2 9JT, United Kingdom*
[2]*Department of Electrical Engineering, King Saud University, Riyadh, Kingdom of Saudi Arabia*

## Summary

Optical wireless communication (OWC) is a promising technology that can provide high data rates while supporting multiple users. The Optical Wireless (OW) physical layer has been researched extensively, however less work was devoted to multiple access and how the OW front end is connected to the network. In this paper, an OWC system which employs a wavelength division multiple access (WDMA) scheme is studied, for the purpose of supporting multiple users. In addition, a cloud/fog architecture is proposed for the first time for OWC to provide processing capabilities. The cloud/fog-integrated architecture uses visible indoor light to create high data rate connections with potential mobile nodes. These optical wireless nodes are further clustered and used as fog mini servers to provide processing services through the optical wireless channel for other users. Additional fog processing units are located in the room, the building, the campus and at the metro level. Further processing capabilities are provided by remote cloud sites. A mixed-integer linear programming (MILP) model was developed and utilised to optimise resource allocation in the indoor OWC system. A second MILP model was developed to optimise the placement of processing tasks in the different fog and cloud nodes available. The optimisation of tasks placement in the cloud-/fog-integrated architecture was analysed using the MILP models. Multiple scenarios were considered where the mobile node locations were varied in the room and the amount of processing and data rate requested by each optical wireless node is varied. The results help identify the optimum colour and access point to use for communication for a given mobile node location and OWC system configuration, the optimum location to place processing and the impact of the network architecture. Areas for future work are identified.

## 1. Introduction

Since the number of wireless communication users has increased dramatically, the demand for wider bandwidths and higher data rates has grown. Broadband radio frequency (RF) technology— the current wireless technology widely utilised in indoor environments—has a number of limitations. The limited available radio spectrum is one of these limitations, which may cause limited channel capacity and low transmission rates. As a result, diverse efficient techniques—such as smart antennas, advanced modulation, and multiple input and multiple output (MIMO) systems have been proposed to improve the use of the radio spectrum and to overcome these limitations [1, 2]. However, achieving data rates above 10 Gbps, for each user, is challenging when using the congested radio spectrum. By 2021, Cisco expects that the Internet traffic will increase 27 times [3]; thus, the increasing demand for high data rates is driving researchers to seek alternative parts of the spectrum, other than the 300GHz of radio spectrum currently in use and proposed for near future use. The optical spectrum is a potential solution. It offers excellent channel characteristics indoor, abundant bandwidth and established low cost components [4 – 11]. Recently, many studies have shown that video, data, and voice can be transmitted through OWC systems at high data rates of up to 25 Gbps and beyond in indoor environments [12–19].

The multipath nature of OWC channels, the limited modulation bandwidth of light sources, and the need to support multiple users, are three drivers of the need for OWC multiplexing and multiple access techniques. Different configurations of transmitters and receivers, in terms of their number and directionality, have been shown to help in reducing the delay spread and increasing the signal-to-noise ratio [20 – 28]. To avoid degradation of the signal quality due to multiple users, efficient utilisation of resources is necessary. Multiplexing of OWC space, time, power, and wavelength resources have thus, recently, attracted the attention of researchers.

With the aim of reducing service latency and power consumption, recent research has focused on proposing new architectures and solutions for cloud computing paradigms. Some of this research has studied distributed computing paradigms that support the applications' high demand and offloading processing off central data centres. In this respect, distributed cloud (DC) is a new generation of cloud computing, where DC tasks are 'distributed' to mini data centre locations, known as fog data centres, resulting in faster services and decreased workload burdens, thereby minimising central data centres' power consumption.

However, for the paradigm to be efficient, distributed applications require a fast communication medium. With data rates of up to 25 Gbps and beyond [8], [9], OWC can satisfy the needs of these distributed applications and can be viewed as a promising fog-based medium for supporting such a paradigm.

This paper proposes an indoor multiple-access OWC network, used in conjunction with cloud/fog-integrated architecture, to create connections with potential mobile nodes, clustered as a fog mini servers to provide processing services. To the best of our knowledge, no previous work has proposed an integration of fog computing with OWC systems. The rest of this paper is organised as follows: Section 2 reviews multiple access techniques in OWC and fog computing and distributed processing. MILP optimisation of resource allocation in optical wireless systems is discussed in Section 3. Section 4 describes an optimum placement of processing to minimise power consumption, and Section 5 presents the conclusions.

# 2. Literature review

## (a) Multiple access schemes in OWC

Multiple access (MA) schemes have been considered for application to OWC systems. Some of the MA schemes used in RF systems are also used in OWC systems. The well-known MA schemes in OWC systems include time division multiple access (TDMA), frequency division multiple access (FDMA), code division multiple access (CDMA), space division multiple access (SDMA), wavelength division multiple access (WDMA) and non-orthogonal multiple access (NOMA).

A TDMA scheme for an OWC system was proposed and investigated in [29]. The scheme is further used to prevent collisions between two transmitters that are equidistant from a receiver [30]. However, the throughput obtained by TDMA schemes decreases when the number of users increases. In addition, both the time and cost increase due to the requirement for synchronisation between the transmitters [31].

FDMA is a scheme that can support multiple access. In OWC, four main techniques based on FDMA have been investigated [32–34]. Single carrier FDMA (SC-FDMA), which is based on frequency division was proposed by the authors in [34] to support users. It reduces the value of the peak-to-average power ratio (PAPR), which is one of its main advantages [35]. The second technique, orthogonal frequency division multiple access (OFDMA), proposed by [32], is a multi-user approach that extends OFDM modulation by allowing each user to be allocated to a group of subcarriers for each time slot. OFDMA has been considered in several studies [36–38] showing that the data rate achieved can be reduced if spectrum partitioning is used [38].

The third technique, orthogonal frequency division multiplexing interleave division multiple access (OFDM-IDMA) was proposed by [32] and compared with OFDMA. OFDM-IDMA is also a multi-user technique based on OFDM modulation. It is an intermediate scheme, situated between OFDM and IDMA technologies. Both OFDMA and OFDM-IDMA are asymmetrically clipped at the zero level after OFDM modulation. OFDM-IDMA provides good results in terms of power efficiency compared to the OFDMA. In addition, the authors in [32] reported that, when the SNR is above 10 dB in a system with a modulation size of 16, the OFDM-IDMA provides better results than OFDMA. However, the decoding complexity and PAPR were found to be lower in OFDMA than in the OFDM-IDMA. The last technique is interleaved frequency division multiple access (IFDMA), which was proposed by [33]. IFDMA was shown to reduce PAPR compared to the OFDMA. In addition, the effects of the non-linear characteristics of LEDs are relatively insignificant in IFDMA, which results in a relatively high power efficiency. The computational complexity is lower in IFDMA than in OFDMA, since IFDMA does not include either discrete Fourier transform or inverse discrete Fourier transform operations. Moreover, the authors in [39] reported that IFDMA can mitigate multi-path distortion and reduce synchronisation errors.

CDMA is a multiple access scheme that can offer a higher spectral efficiency than OFDMA and TDMA. In CDMA, each user, can employ a special code to provide simultaneous transmission and reception. For example, an optical code proposed by [40] is used by each user. In addition, random optical codes have been proposed by [41, 42] for supporting a large number of users by spreading the signal bandwidth. A synchronisation technique was reported by [43] for preventing undesirable correlation characteristics across random optical codes; consequently, good performance was achieved.

Colour shift keying (CSK) was proposed by [44] as a technique, based on the CDMA scheme, for increasing the capacity in multiple access; consequently, each transmitter can achieve 3 dB improvement compared to OOK. Multi-carrier (MC) is another technique based on CDMA, investigated by [45, 46]. MC-CDMA is a combination of the CDMA and OFDM schemes. MC-CDMA diffuses the data symbols of each user in the frequency domain over the OFDM subcarriers. Thereafter, the sum of the data symbols from the multiple users is re-modulated in the time domain of the OFDM scheme. Furthermore, the researchers in [45] stated that the transmitted optical power can be reduced by using sub-carrier selection. Another technique based on CDMA was proposed by [47] and provides interference-free links while transferring information. The technique allows users to decode their signals without pre-confirmation. However, this technique was investigated in a small area of <1 m, which limits its utility.

SDMA is a scheme that can be used in OWC systems and allows multi-user access. In [48], the researchers concluded that, when the number of transmitters is increased, the throughput increases provided interference is avoided. Moreover, it was concluded that the system capacity can be improved at least ten times over, when SDMA is used instead of TDMA. An approach based on SDMA, called low-complexity suboptimal algorithm, for coordinated multi-point OWC system with SDMA grouping, was reported by [49]. The proposed technique offers an improvement in the system performance, throughput, and fairness.

WDMA can support multiple users based on wavelength division multiplexing. It has been studied in OWC systems [7], [50 – 54]. WDMA uses a multiplexer at the transmitter for aggregating different wavelengths from light sources into a single OW beam. Thereafter, at the receiver, a de-multiplexer is used to separate the wavelengths; thus supporting multiple access. The operation of WDMA is similar to that of FDMA scheme; both operating in the frequency domain. The two light sources typically used in OWC systems (LEDs and LDs) were investigated in [53, 54]. When using red, green, and blue (RGB) LEDs, a data rate of more than 3.22 Gbps was achieved in [53] through WDMA implementation in a visible light communication (VLC) system. Moreover, another demonstration of a VLC system that used WDMA was reported in [54]. In addition, by using red, yellow, green, and blue (RYGB) LDs, the researchers in [7] achieved a data rate of up to 10 Gbps.

NOMA, also called power domain multiple access, was studied in OW in [55, 56] as a promising technique for providing multiple access. NOMA differs from other multiple access techniques that provide orthogonal access for multiple users in the frequency, time, code, or phase domains. Each user in NOMA can use the same frequency band at the same time. Therefore, provided a good SINR is achieved through power control, the users can be distinguished. In addition, the transmitter in NOMA applies superposition coding to simplify the operation at the receiver. Consequently, the channels are separated, at the receiver, between the uplink and the downlink users [56]. NOMA was shown in recent studies to support multiple users and provide a high data rate [56 – 60]. NOMA can outperform OFDMA in terms of the data rate in OWC systems, as reported by [56]. In addition, in [59], the researchers reported that the system capacity was improved by using NOMA. In [60], the author reported that the combined application of the MIMO and NOMA can offer a high data rate, high capacity, and high spectral efficiency. An experiment was conducted in [60] that used NOMA, along with MIMO, in a VLC system. A normalised gain difference power allocation (NGDPA) method was proposed to provide efficient low-complexity power allocation. The experimental result showed that the application of NOMA with NGDPA resulted in a sum rate improvement of up to 29%, compared to that of NOMA.

## (b) Fog computing and distributed processing

The vast expansion in the usage of cloud services, and the significant increase in distributed services, call for new architectures and solutions to provision those services at high date rates to the end user while reducing latency and power consumption. Distributed computing has become a popular solution that shifts the workload from the central cloud to the fog and, thus, closer to end-users; hence, distributed cloud technology was developed to provide access to computational resources at the edge of the network, in close proximity to the end user, instead of accessing the central cloud. This has resulted in faster services and lower computing burdens, thereby minimising central data centre power consumption, and reducing the overall power consumption and latency. Proposed and tested architectures that accommodate the concept and features of distributed cloud are referred to as 'cloudlets' [61], 'fog' [62], or 'edge computing' [63]. Processing at the edge nodes, and the networking fabric used to enable this, are considered to be major attributes of fog architectures. These attributes play an important role in delivering a service with lower latency and power consumption.

The type of edge nodes affects the performance of the processing task and, therefore, the delivery of the requested services. Building small distributed data centres, using a smaller number of high performance servers, [64], has been proven to deliver good throughput with lower power consumption and latency. A newer approach aims to build a cluster of fog nodes out of underutilised computing resources in computer clusters [65]. This approach has led to a new definition of the type of fog nodes that can be used as processing nodes. With the huge expansion in the number of smart devices and IoT, fog can comprise any smart devices clustered as a single fog mini data centre to provide processing services. Such a paradigm is built from available idle or underutilised resources that become available opportunistically. This leads to new fog frameworks comprised of IoT nodes [66], vehicles [67], mobile phones [68], and any other portable devices [69].

The networking fabric also plays an important role in fog-based architectures. Fabrics with high data-rate communication enhances the connection between the end user and fog nodes, and thereby achieves improved service delivery. Communication technologies that can support fog-based computing can be found in [70]. Optical communication has been shown to satisfy the required data rate for distributed mini data centres [71]. Since most of the edge smart devices are equipped with wireless/cellular networking capabilities, these are the two main mediums currently supporting end user connections to opportunistic fog nodes. However, with the recent huge expansion of services and application demands, these mediums may not support the required data rates needed. This calls for more research into the application of high data-rate mediums, such as wired fibre-optics, wired visible light, wireless visible light communication and optical wireless communication in fog-based networks. This paper provides the first study to the best of our knowledge where optical wireless is considered and integrated with opportunistic and fixed fog considering the optical wireless resource allocation mechanisms needed and the optimum placement of processing jobs all the way from the optical wireless handset / mobile unit to the central cloud passing through different fixed fog options at the room, building, campus and metro levels.

# 3. MILP optimisation for resource allocation in optical wireless

Wavelength division multiplexing (WDM) can be used in both the uplink and downlink of OWC systems. In VLC systems, wavelengths in the visible light range can be optically summed to form the white light used in illumination and communication, as in [72] where four colours: RYGB LDs were used. In the uplink, to avoid the problem of glare, infra uplink design was proposed in [73, 74]. In this study, only the downlink will be considered. We assumed a room with dimensions as shown in Table 1, in which the parameters of the transmitter and receiver are also given.

**Table 1.** Room Configurations

| Parameters | Configurations | |
|---|---|---|
| Room | | |
| Length × Width × Height | 8 m × 4 m × 3 m | |
| Walls and Ceiling reflection coefficient | 0.8 | |
| Floor reflection coefficient | 0.3 | |
| Number of Reflections | 1 | 2 |
| Area of reflection element | 5 cm × 5 cm | 20 cm × 20 cm |
| Order of Lambertian pattern, walls, floor and ceiling | 1 | |
| Semi-angle of reflection element at half power | $60^0$ | |
| Transmitters | | |
| Number of transmitters' units | 8 | |
| Transmitters locations (x, y, z) | (1 m, 1 m, 3 m), (1 m, 3 m, 3 m), (1 m, 5 m, 3 m), (1 m, 7 m, 3 m), (3 m, 1 m, 3 m), (3 m, 3 m, 3 m), (3 m, 5 m, 3 m) and (3 m, 7 m, 3 m) | |
| Number of RYGB LDs per unit | 9 | |
| Transmitted optical power of Red LD | 0.8 W | |
| Transmitted optical power of Yellow LD | 0.5 W | |
| Transmitted optical power of Green LD | 0.3 W | |
| Transmitted optical power of Blue LD | 0.3 W | |
| Total transmitted power of RYGB LD | 1.9 W | |
| Semi-angle at half power | $60^0$ | |
| Receiver | | |
| Number of Photodetectors | 1 | |
| Area of the photodetector | 1 $cm^2$ | |
| Responsivity | 0.4 A/W | |
| Field of view (FOV) | $40^{\circ\,0}$ | |
| Receiver noise current spectral density | 4.47 pA/√Hz [76] | |
| Receiver bandwidth | 5 GHz [76] | |

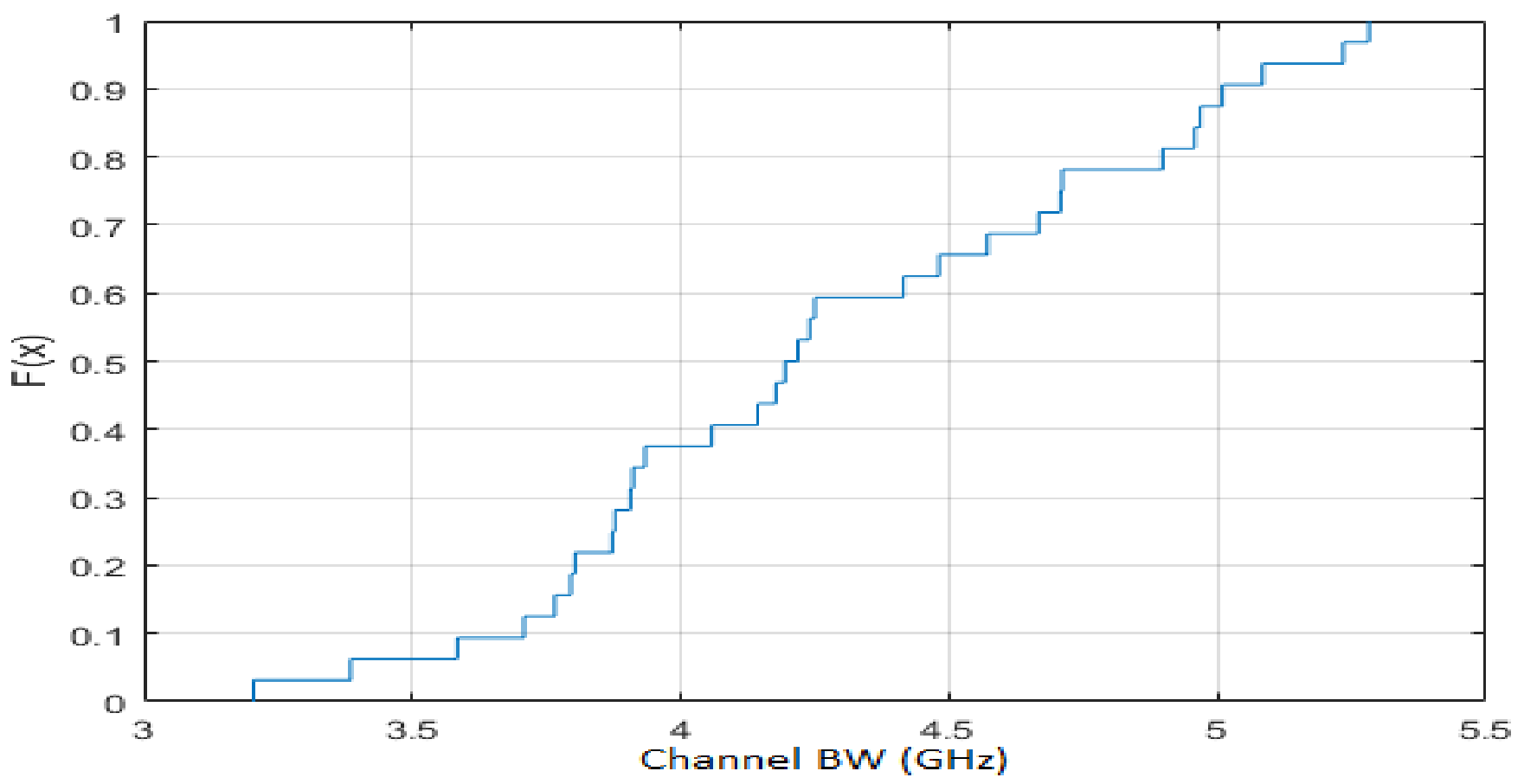

**Figure 1.** CDF of the optical channel bandwidth at different locations in the room using the parameters in Table 1.

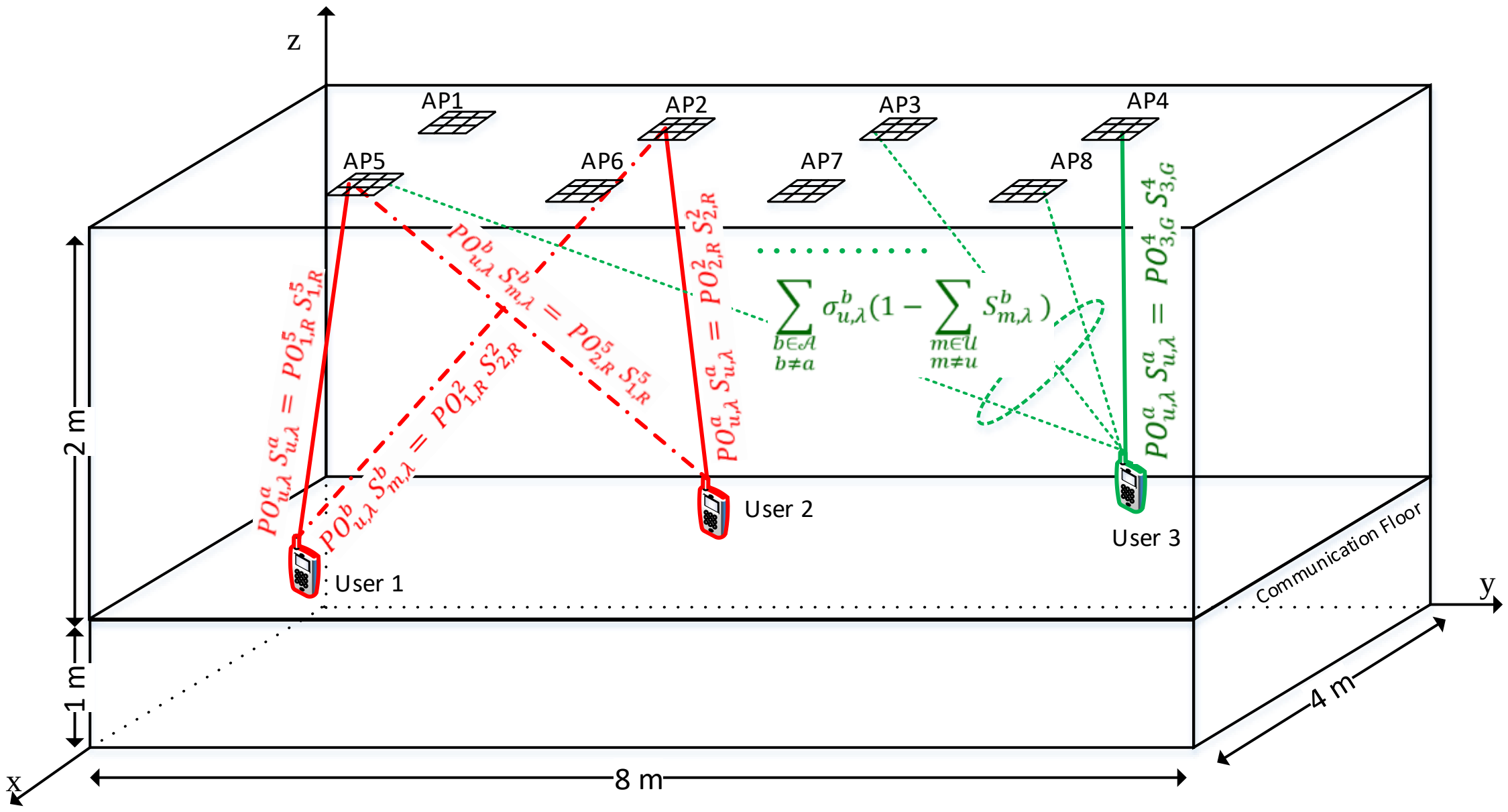

**Figure 2.** A room with three users. Solid lines indicate assignment of an access point to a user. Dot-dashed lines show interference between users using the same wavelength. Dotted lines indicate unmodulated wavelengths from access points causing background light shot noise.

Users were distributed over this room using a 2D Poisson point process (PPP) and two 8-users scenarios with fixed user locations were considered. The channel was characterised using a simulation package similar to that in [75]. Up to the second reflections were considered in this work as higher order reflections have no significant impact on the received power [75]. The channel impulse response was then used to calculate the delay spread and the optical channel bandwidth using the parameters in Table 1.

The cumulative distribution function for the optical channel bandwidth composed using 128 locations in the room is shown in Figure 1. It can be seen that around 60% of the locations in the room support a bandwidth of at least 4 GHz. This was achieved by reducing the FOV of the receiver to limit the number of input rays which results in decreasing the delay spread and hence increasing the bandwidth.

A MILP model was developed to optimise WDM wavelength assignment to maximise the sum of SINRs for all users [77]. The precalculated values of the channel impulse response were used to calculate the optical power received at each potential location in the room from each access point (AP) using different wavelengths. The MILP model was then used to assign APs and wavelengths to users so that the sum of SINRs was maximised, based on optical channel information. Figure 2 illustrates a scenario with three users. Users 1 and 2 suffer from background light shot noise in addition to interference as they are assigned the red wavelength, while user 3 suffers only from background light shot noise as the green wavelength is not assigned to other users. Before introducing the MILP model, we define the sets, parameters, and variables used, and describe how the SINR is calculated;

*Sets:*

| | |
|---|---|
| $\mathcal{U}$ | Set of users in the room; |
| $\mathcal{A}$ | Set of access points; |
| $\mathcal{W}$ | Set of available wavelengths (RYGB); |

*Parameters:*

| | |
|---|---|
| $u$, $m$ | User indices; $u$ is desired user, $m$ indexes other users who may cause interference; |
| $a$, $b$ | Access point indices; $a$ is the access point allocated to user $u$, $b$ is an access point allocated to another user $m$; |
| $\lambda$ | Wavelength index; |
| $PO^{a}_{u,\lambda}$ | Optical power received by user $u$ from an access point $a$ using wavelength $\lambda$. This value was precalculated using a channel modeling tool, where the LOS and first order reflection components were calculated for the given access point and user location and for the given wavelengths; |
| $P^{a}_{u,\lambda}$ | The squared electrical current at the receiver of user $u$ due to the optical power received from access point $a$ at the wavelength $\lambda$. (Note that the squared current and electrical power are equivalent for a given system input impedence); |
| $\sigma^{b}_{u,\lambda}$ | The shot noise mean square current at the receiver of user $u$ due to the background unmodulated power of access point $b$ operating at wavelength $\lambda$; |
| $\sigma_{Rx}$ | The mean square receiver noise current; |
| $R^{a}_{u}$ | The data rate supported based on optical channel bandwidth for the channel between access point $a$ and user $u$ if OOK modulation is used. This value was precalculated using the channel modeling tool for the given access point and the given user. |

Variables:

| | |
|---|---|
| $\gamma^{a}_{u,\lambda}$ | SINR of user $u$ assigned to access point $a$ and wavelength $\lambda$; |
| $S^{a}_{u,\lambda}$ | A selector function where a binary value of 1 indicates the assignement of user $u$ to access point $a$ and wavelength $\lambda$ (see Figure 2); |
| $\phi^{a,b}_{m,\lambda,u}$ | Non-negative linearisation variable, where $\phi^{a,b}_{m,\lambda,u} = \gamma^{a}_{u,\lambda}\, S^{b}_{m,\lambda}$; |

To calculate the SINR, different powers are calculated as follows:

The electrical signal power received by user $u$ from access point $a$ and using wavelength $\lambda$ is calclulated as:

$$P^{a}_{u,\lambda} = (R\, PO^{a}_{u,\lambda}\, h^{a}_{u,\lambda})^{2} \quad (1)$$

where $R$ is the responsivity of the photodetector in (A/W) and $h^{a}_{u,\lambda}$ is the DC channel gain between access point $a$ and user $u$ for wavelength $\lambda$.

The preamplifier noise is given by:

$$\sigma_{Rx} = N_{pr} B \quad (2)$$

where $N_{pr}$ is the preamplifier noise power density in ($A^{2}$/Hz) .

The background light shot noise is calculated as:

$$\sigma^{b}_{u,\lambda} = 2e\left(R\, PO^{b}_{u,\lambda}\, h^{b}_{u,\lambda}\right)B \quad (3)$$

where $e$ is the electron charge (C) and $B$ is the bandwidth. The SINR of user $u$, who is assigned wavelength $\lambda$ of access point $a$ is therefore expressed as:

$$\gamma_{u,\lambda}^{a} = \frac{Signal}{Interference + Noise} = \frac{P_{u,\lambda}^{a}\, S_{u,\lambda}^{a}}{\sum_{\substack{b\in\mathcal{A}\\ b\neq a}} \sum_{\substack{m\in\mathcal{U}\\ m\neq u}} P_{u,\lambda}^{b}\, S_{m,\lambda}^{b} + \sum_{\substack{b\in\mathcal{A}\\ b\neq a}} \sigma_{u,\lambda}^{b} \left[1 - \sum_{\substack{m\in\mathcal{U}\\ m\neq u}} S_{m,\lambda}^{b}\right] + \sigma_{Rx}} \tag{4}$$

where $S_{u,\lambda}^{a}$ is a binary assignment variable that is equal to 1 if user $u$ is assigned to access point $a$ and wavelength $\lambda$. The first term in the denominator is the interference, which was calculated by summing the power received by user $u$ from all APs where the same wavelength was used for communication, but assigned to other users, hence, causing interference. Furthermore, noise was calculated by summing two terms representing the receiver noise $\sigma_{Rx}$ (calculated using Equation 2) which is constant for all users with identical receivers and background light shot noise. The background light shot noise $\sigma_{u,\lambda}^{b}$ is calculated using (3) which is the power received by the current user from all APs emitting unmodulated wavelengths identical to the current user's assigned wavelength (used for illumination only), hence causing shot noise. In other words, interference was calculated by summing the signal powers of modulated light beams of the same wavelength, while background light shot noise was calculated by summing the signal powers of unmodulated wavelengths.

The interference term was calculated as the sum of the squared electrial currents and not by squaring the sum of the received optical powers multiplied by the responsivity of the photodetector. This simplifies the MILP implementation by maintaing linearity inside the MILP, while the squaring is carried out outside the MILP, (pre-calculated). The error due to this method of calculation can be reduced by using the Cauchy-Schwarz inequality and introducing a factor $n$ for the number of interferers which can be either 0, 1, or 2 due to the tightened receiver's FOV. In the first two cases, there will be no error in the calculations. The error is only encountered when there are two interfering access points. Shot noise due to the interfering signal power was also ignored as it is (for our system parameters) about four orders of magnitude lower than the signal power for 1 μW of received optical power.

Rewriting equation (4):

$$\sum_{\substack{b\in\mathcal{A}\\ b\neq a}} \sum_{\substack{m\in\mathcal{U}\\ m\neq u}} \gamma_{u,\lambda}^{a}\, P_{u,\lambda}^{b}\, S_{m,\lambda}^{b} + \sum_{\substack{b\in\mathcal{A}\\ b\neq a}} \gamma_{u,\lambda}^{a} \sigma_{u,\lambda}^{b} \left[1 - \sum_{\substack{m\in\mathcal{U}\\ m\neq u}} S_{m,\lambda}^{b}\right] + \gamma_{u,\lambda}^{a}\, \sigma_{Rx} = P_{u,\lambda}^{a}\, S_{u,\lambda}^{a} \qquad \forall u \in \mathcal{U}, \forall a \in \mathcal{A}, \forall \lambda \in \mathcal{W} \tag{5}$$

which can be rearranged as:

$$\sum_{\substack{b\in\mathcal{A}\\ b\neq a}} \sum_{\substack{m\in\mathcal{U}\\ m\neq u}} \left[P_{u,\lambda}^{b} - \sigma_{u,\lambda}^{b}\right] \gamma_{u,\lambda}^{a}\, S_{m,\lambda}^{b} + \sum_{\substack{b\in\mathcal{A}\\ b\neq a}} \gamma_{u,\lambda}^{a} \sigma_{u,\lambda}^{b} + \gamma_{u,\lambda}^{a}\, \sigma_{Rx} = P_{u,\lambda}^{a}\, S_{u,\lambda}^{a}. \tag{6}$$

The first term containing the interference and background light shot noise from other access points is a nonlinear quadratic term involving the multiplication of a continuous variable by a binary variable. Linearisation was performed following the same procedure in [78] as shown in equations (11)–(14) below.

The MILP model is defined as follows:

Objective: Maximise the sum of SINRs for all users,

$$Maximise \sum_{a\in\mathcal{A}} \sum_{u\in\mathcal{U}} \sum_{\lambda\in\mathcal{W}} \gamma_{u,\lambda}^{a} \tag{7}$$

Subject to:

$$\sum_{u \in \mathcal{U}} S^a_{u,\lambda} \leq 1 \qquad \forall a \in \mathcal{A}, \forall \lambda \in \mathcal{W} \qquad (8)$$

Constraint (8) ensures that a wavelength belonging to an AP is only allocated once

$$\sum_{a \in \mathcal{A}} \sum_{\lambda \in \mathcal{W}} S^a_{u,\lambda} \geq 1 \qquad \forall u \in \mathcal{U} \qquad (9)$$

$$\sum_{a \in \mathcal{A}} \sum_{\lambda \in \mathcal{W}} S^a_{u,\lambda} \leq 1 \qquad \forall u \in \mathcal{U} \qquad (10)$$

Constraints (9) and (10) ensure that a user is assigned one wavelength only. The following constraints (11)–(14) were used to linearise the multiplication process of the continuous variable by the binary variable in the quadretic term, where the non-negative linearisation variable $\phi^{a,b}_{m,\lambda,u} = \gamma^a_{u,\lambda} S^b_{m,\lambda}$ is introduced:

$$\phi^{a,b}_{m,\lambda,u} \geq 0. \qquad (11)$$

$$\phi^{a,b}_{m,\lambda,u} \leq \beta S^b_{m,\lambda} \qquad \forall u, m \in \mathcal{U}, \forall a, b \in \mathcal{A}, \forall \lambda \in \mathcal{W} \quad (u \neq \mathrm{m}, a \neq b) \qquad (12)$$

where $\beta$ is a large number, so that $\beta \gg \gamma$.

$$\phi^{a,b}_{m,\lambda,u} \leq \gamma^a_{u,\lambda} \qquad \forall u, m \in \mathcal{U}, \forall a, b \in \mathcal{A}, \forall \lambda \in \mathcal{W} \quad (u \neq \mathrm{m}, a \neq b) \qquad (13)$$

$$\phi^{a,b}_{m,\lambda,u} \geq \beta S^b_{m,\lambda} + \gamma^a_{u,\lambda} - \beta \qquad \forall u, m \in \mathcal{U}, \forall a, b \in \mathcal{A}, \forall \lambda \in \mathcal{W} \quad (u \neq \mathrm{m}, a \neq b) \qquad (14)$$

Using this linearisation variable to replace the quadratic term, Equation (6) can be re-written as:

$$\sum_{\substack{b \in \mathcal{A} \\ b \neq a}} \sum_{\substack{m \in \mathcal{U} \\ m \neq u}} [P^b_{u,\lambda} - \sigma^b_{u,\lambda}] \, \phi^{a,b}_{m,\lambda,u} + \sum_{\substack{b \in \mathcal{A} \\ b \neq a}} \gamma^a_{u,\lambda} \sigma^b_{u,\lambda} + \gamma^a_{u,\lambda} N_0 = P^a_{u,\lambda} S^a_{u,\lambda}. \qquad (15)$$

In order to support a BER of $10^{-9}$ using OOK modulation (our chosen modulation format here), the SINR should not go below 15.6 dB. The same performance (BER of $10^{-9}$) can however be achieved with a lower SINR using forward error correction (FEC) techniques at the expense of increased data rate overhead[1]. Here 10% overhead is assumed when SINR decreases to 14 dB.This is added as a constraint:

$$\gamma^a_{u,\lambda} \geq 10^{14.0/10} \qquad \forall u \in \mathcal{U}, \forall a \in \mathcal{A}, \forall \lambda \in \mathcal{W} \qquad (16)$$

The OW system is backhauled and is connected to the fibre access network. Therefore an additional constraint is added to ensure the capacity of the ONU nodes (to which the OW access points are connected), is not exceeded. The capacity of the ONU used in this architeture is $C^{ONU}$=10 Gbps (see Table 4), hence:

$$\sum_{u \in \mathcal{U}} \sum_{\lambda \in \mathcal{W}} S^a_{u,\lambda} R^a_u \geq C^{ONU} \qquad \forall a \in \mathcal{A} \qquad (17)$$

where $R^a_u$ is the data rate supported using the optical channel between access point $a$ and user $u$.

After the assignment of APs and wavelengths was optimised using the MILP model, the optical channel bandwidth, the SINR and achievable data rates were calculated for each user and scenario (see Table 2) in this study and they are shown in Figures (3-5).

It should be noted that the supported data rate can be limited by one of three factors: the modulation bandwidth of the light source, the optical channel bandwidth, or the receiver bandwidth. Since LDs were used in this study, they do not limit the data rate as they can support modulation rates at GHz rates beyond those in our study [79]. To address the second limiting factor, the use of 40° FOV improves the optical channel bandwidth as it limits the number of incident rays from different refelecting elements, hence, the delay spread is reduced and the channel bandwidth increases. This FOV also ensure good room illumination following the approach in [8]. The receiver bandwidth determines the amount of noise addmited from background ligth in the form of shot noise and it also determines the preamplifier noise (Equations 2-3), so reducing the receiver bandwidth leads to an increase in the SINR at the cost of reducing the supported data rate. SINR is also affected by interference which increases as the number of users sharing the same wavelength increases in WDM and also as the amount of overlap between the coverage area of different access points increases. Since VLC is considered in this study, this dictates a limit on the number of wavelengths used, ie four (RYGB) wavelengths; and also dictates the large large coverage area per access point to meet the illumination standards, with an access point's half power semi-angle of 60°. The VLC system performance can however be improved by reducing the FOV of the receiver and by optimizing the allocation of access points and wavelength resources to minimise interference so that the overall sum SINR is maximised.

[1] For more information refer to http://www.ieee802.org/3/10G_study/public/july99/azadet_1_0799.pdf where for example it is stated that "RS(255,239) overhead is 6% for input BER=$10^{-4}$, output BER=$10^{-14}$"

The calculated values of data rates for different users are to be used in the next section where a MILP model is developed to optimise the placement of the processing in the integrated cloud/fog with OWC to minimise the overall power consumption.

The results in Figs. 3, 4 and 5 show that all users achieve an SINR above our required threshold of 14 dB with per user data rates above 3 Gb/s and below 7 Gb/s for our system parameters.

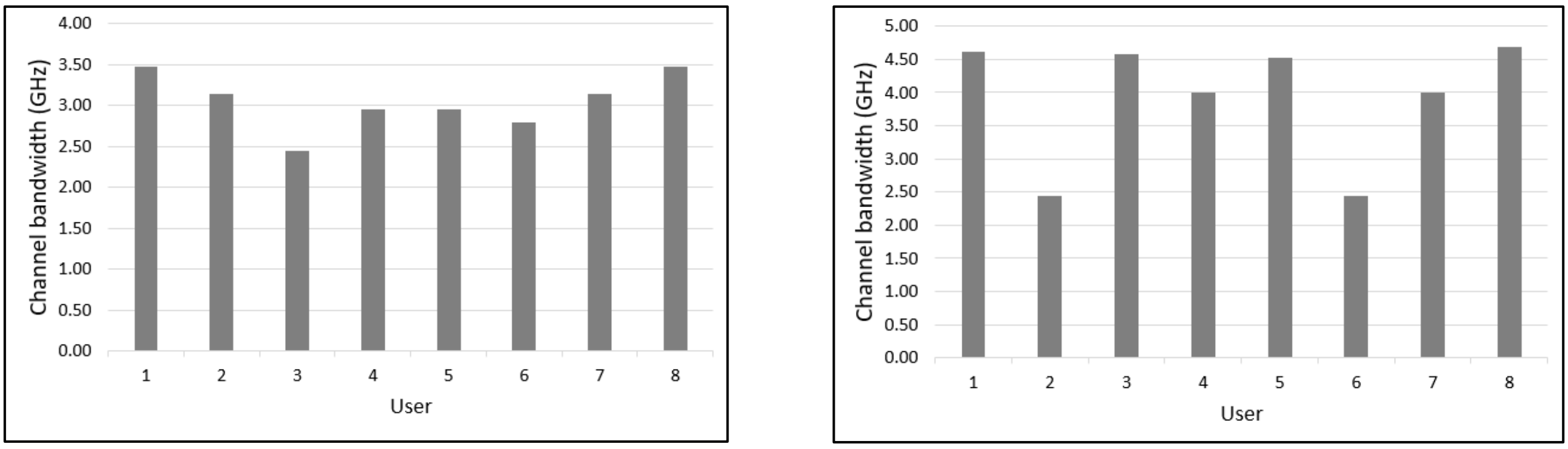

(a) Scenario 1 (b) Scenario 2

**Figure 3.** Optical channel bandwidth.

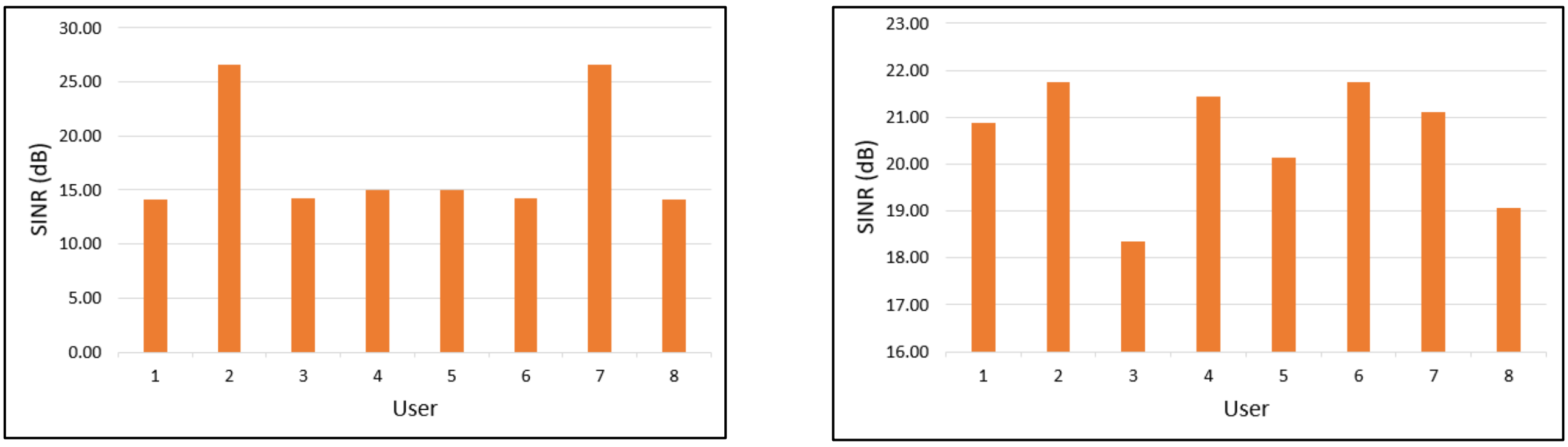

(a) Scenario 1 (b) Scenario 2

**Figure 4.** SINR for different users.

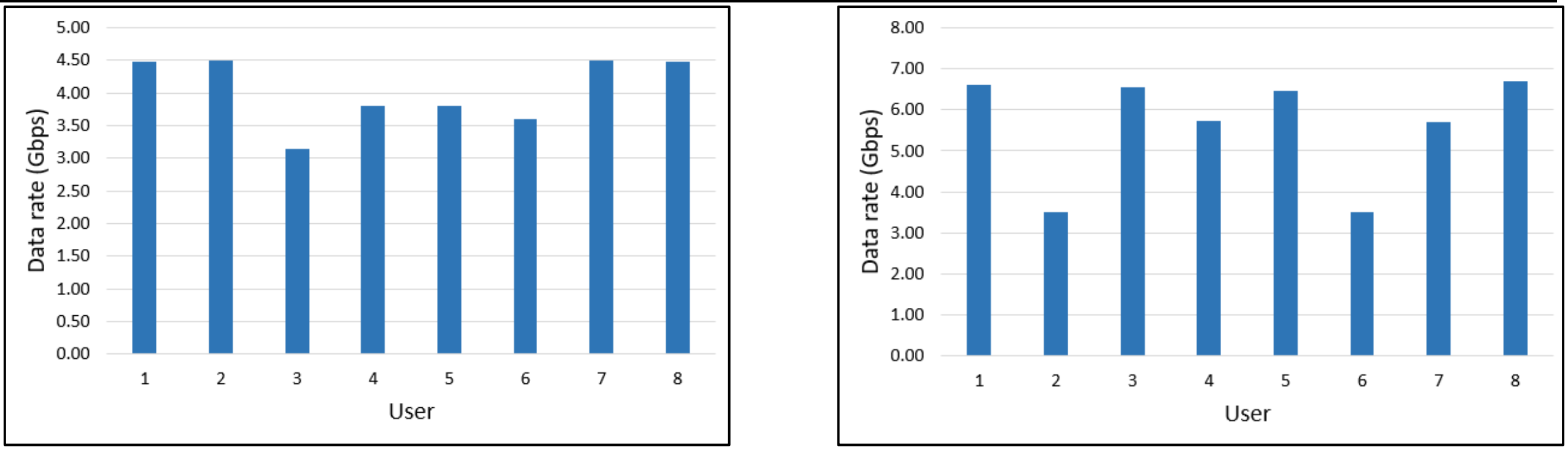

(a) Scenario 1 (b) Scenario 2

**Figure 5.** Data rates for different users.

# 4. Optimum placement of processing to minimise power consumption

The proposed integrated cloud/fog architecture is shown in Figure 6 which builds on our work in the optimisation of distributed data centres [80 - 83], network architecture optimisation [84 – 88] and energy efficient routing [89 – 93]. It consists of one or more OW mobile users in a room clustered as a mobile fog unit (MobFog). Each mobile device communicates with one or more light units (access points) and is assigned to one of the channel wavelengths, Red, Yellow, Green, or Blue (RYGB). All access points are connected to a passive optical network (PON), where each access point is also connected to an optical network unit (ONU). All ONUs are connected to a central optical line terminal (OLT) located in the same room. The room is also equipped with a commodity server (low-end computer), which acts as a mini fog node (RoomFog). This node is connected to the central OLT through an ONU and an optical link. The proposed architecture introduces three more fog data centres located in the building (BuildFog), campus (CampFog) and metro network layer (MetroFog). This architecture is integrated with the central cloud data centre (CCloud) through an optical infrastructure to support high demand requests that cannot be fulfilled by mobile units or fog nodes.

The OLT acts as a controller unit [94] that collects processing requests from mobile units present in the same room. These requests can be generated from applications in the mobile units. The mobile units transmit the data to be processed. The knowledge extracted after processing the data (for example (i) the presence of absence of someone in a transmitted video sequence; or (ii) whether a person is fine or not after transmitting a large heart rate signal) is always smaller than the transmitted data [95–97]. The OLT also assigns in an optimal way (MILP in this section) the collected requests to the Central Cloud or any fog node to be processed. When the OLT decides that a task / demand is to be processed by the MobFog, it allocates this task to the participants' mobiles through VLC communication and then forwards the processing results back to the OLT. If the OLT decides to assign the demand to the RoomFog, BuildFog or CampFog nodes, the demand will be sent through a local Ethernet LAN to the required location. For the MetroFog and CCloud assignment, demands traverse through the optical infrastructure to either location. The tasks assignment was optimised using a MILP model to minimise the power consumption of the overall architecture while considering different optical wireless data rates depending on the scenario. Below are the notation of the parameters and the variables used in this optimisation model.

In this MILP model, the sets are defined as follows:

| | |
|---|---|
| $N$ | Set of all nodes; |
| $Nm_i$ | Set of nodes that are neighbours of node *i;* |
| $PN$ | Set of processing nodes; |
| $SN$ | Set of source nodes; |
| $K$ | Set of tasks (demands); |

The following parameters are also defined:

| | |
|---|---|
| $W_{ks}$ | Workload demand of task $k$ generated from source node $s$, in million instructions per second (MIPS). |
| $F_{ks}$ | Flow (data rate) demand of task $k$ generated from source node $s$ (in Mbps). |
| $C_n$ | Workload capacity of processing node $n$ (in MIPS). |
| $L_{ij}$ | Capacity of the link between nodes $i$ and $j$ (in Mbps). |
| $E_n$ | Power per MIPS of processing node $n$ (W/MIPS). |
| $\Psi_n$ | Power per Mbps of the route to the processing node $n$ (W/Mbps). |

The following variables are also defined:

| | |
|---|---|
| $\mathrm{P}_n$ | Total power consumption due to data processing. |
| $\mathcal{P}_n$ | Total power consumption due to networking. |
| $\mathrm{X}_{kn}$ | Task $k$ processing workload, in MIPS, assigned to processing node $n$. |
| $\delta_{kn}$ | Binary variable, $\delta_{kn} = 1$ if task $k$ is assigned to processing node $n$, 0 otherwise. |
| $L_{ksd}$ | Traffic flow of task $k$ sent from source node $s$ to processing node $d$. |
| $\lambda_{ij}^{ksd}$ | Traffic flow of task $k$ sent from source node $s$ to destination (processing node) $d$ through physical link $i$ and $j$. |

The model's objective is to minimise the power consumption of the overall proposed architecture, including the processing and networking power consumption of the processing locations, and the network connecting these locations, as presented below in Equation (18).

Objective: Minimise

$$\sum_{n \in PN} P_n + \sum_{n \in PN} \mathcal{P}_n . \tag{18}$$

The objective equation consists of two terms: the power consumed by processors due to computation and the power consumed by transmitting traffic through the network. Note that in our network topology, in Figure. 6, there is a single route between the OW mobile units and each processing node option (CCloud, MetroFog, CampFog, BuildFog, RoomFog, or other mobile units in the MobFog). The power consumption of this route in W/Mbps is given by $\Psi_n$ in Equation (20). Therefore, the second term in Equation (18) summed over processing nodes (as there is a single network path to each processing node in our case).

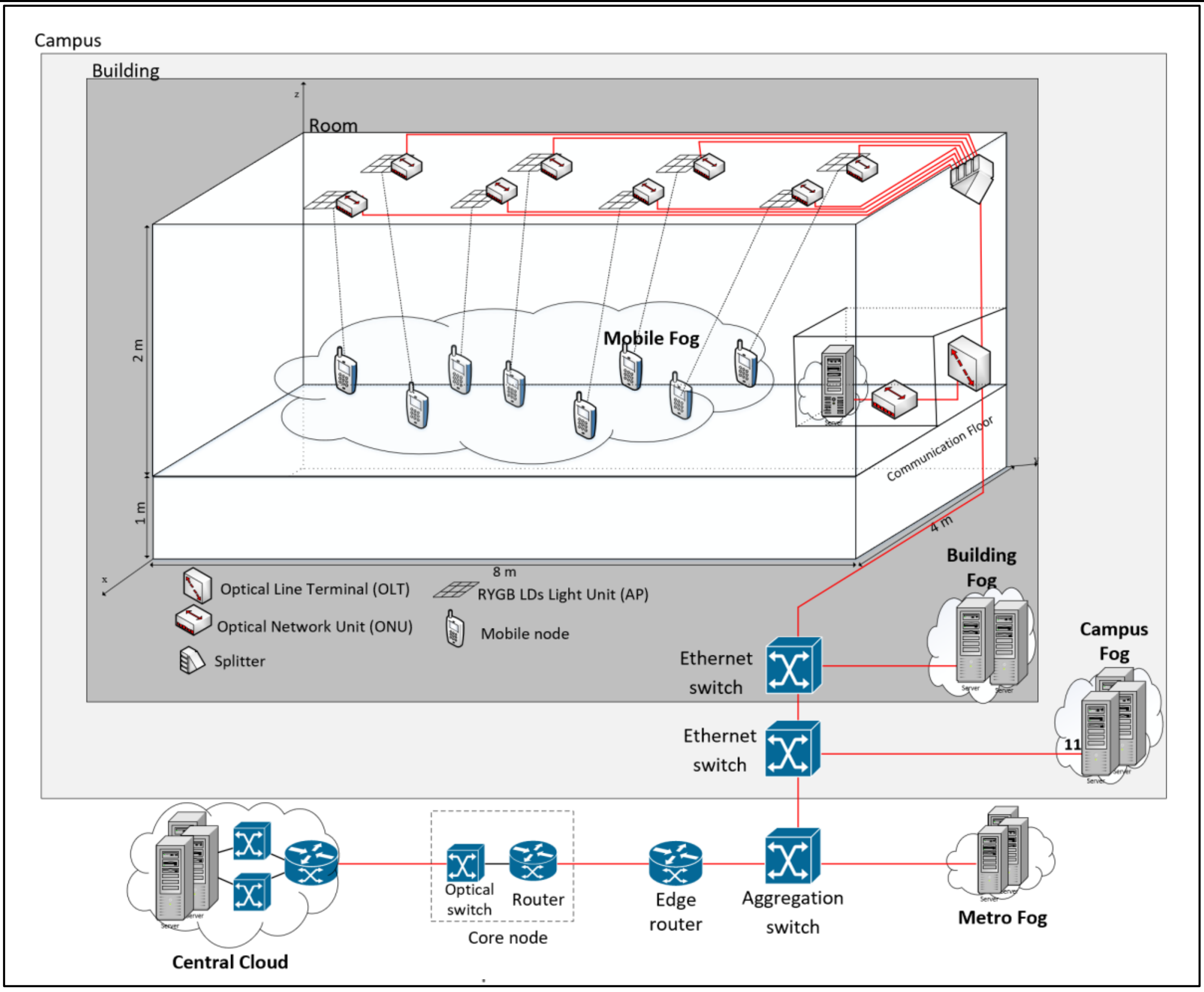

**Figure 6.** Cloud/fog-based architecture.

The processing power consumption, $P_n$ is given as:

$$P_n = \sum_{k \in K} X_{kn} \quad E_n \qquad \forall \ n \in PN, \tag{19}$$

where $X_{kn}$ is the workload demanded by task $k$, in million instructions per second (MIPS), assigned to processing node $n$. $E_n$ is the energy in watts per MIPS of the node processor, calculated using the maximum processing capacity of the node.

The networking power consumption, $\mathcal{P}_n$ is given as:

$$\mathcal{P}_n = \sum_{k \in K} \delta_{kn} \quad F_{ks} \ \Psi_n \qquad \forall \ n \in PN, \ \ s \in SN, \tag{20}$$

where $\delta_{kn}$ is a binary variable which specifies the assignment of task $k$ to processing node $n$, and $F_{ks}$ is the task data rate demand (in Mbps) generated from source node $s$. $\Psi_n$ is the total power per Mbps of all nodes between the source and the assigned processing node (in Watt per Mbps). For the optical wireless link, the power per Mbps value is calculated based on the individual wavelength colour (RYGB) used for the connection.

The model is subject to the following constraints:

Processing allocation constraints:

$$\alpha \ X_{kn} \ \geq \ \delta_{kn} \qquad \forall \ k \in K, n \in PN \tag{21}$$

$$X_{kn} \ \leq \ \alpha \ \delta_{kn} \qquad \forall \ k \in K, n \in PN \tag{22}$$

Constraints (21) and (22) ensure that task $k$ is assigned to processing node $n$.

$$\sum_{n \in PN} \delta_{kn} \ = \ 1 \qquad \forall \ k \in K \tag{23}$$

Constraint (23) ensures that each task *k* will be assigned to one processing node.

Processing node capacity constraint:

$$\sum_{k \in K} X_{kn} \leq C_n \qquad \forall\ n \in PN \tag{24}$$

Constraint (24) ensures that each task *k* assigned to a processing node *n* does not exceed the processing capacity of this processing node.

Link capacity constraint:

$$\sum_{k \in K} \sum_{\substack{s \in SN \\ d \in PN}} \lambda_{ij}^{ksd} \leq L_{ij} \qquad \forall\ i \in N, j \in Nm_i, i \neq j \tag{25}$$

Constraint (25), ensures that the traffic of task *k* sent from source *s* to processing node *d* does not exceed the capacity of the link between any two nodes *i* and *j*.

Flow conservation constraint:

$$\sum_{\substack{j \in Nmi \\ i \neq j}} \lambda_{ij}^{ksd} - \sum_{\substack{j \in Nmi \\ i \neq j}} \lambda_{ji}^{ksd} = \begin{cases} L_{ksd} & if\ i = s \\ -L_{ksd} & if\ i = d \\ 0 & otherwise \end{cases} \qquad \forall\ k \in K, s \in SN, d \in PN, i, j \in N \tag{26}$$

Constraint (26), ensures that the total incoming traffic is equal to the total outgoing traffic for all nodes except for the source and destination nodes.

$$L_{ksd} = F_{ks}\ \delta_{kd} \qquad \forall\ k \in K, s \in SN, d \in PN. \tag{27}$$

Constraint (27), ensures that the traffic from source node *s* to destination node *d* is equal to the data rate of task *k* generated from source *s*. $\delta_{kd}$ is a binary variable used to ensure that task *k* is assigned to destination *d*.

The main outputs of the MILP model in this section are the values of the $\delta_{kd}$ decision variable. Therefore, this is an optimal placement and routing problem solution. Based on the value of $\delta_{kd}$, the networking and processing power consumption splits can be determined.

The flow process of any generated request follows six phases. Firstly, a request is generated by a mobile unit and sent to the OLT, which has full knowledge of the available resources. Secondly, the OLT sends a positive acknowledgment to the source node. These two phases are not considered in the model as they generate negligible traffic. Thirdly, the data to be processed are sent in an uplink from the mobile source node to the OLT through the optical wireless channel. This phase is also not considered in the model as it is a common phase for all requests and will not affect the placement decision. Fourthly, the data to be processed are offloaded from the OLT to one of the available processing nodes (MobFog, RoomFog, CampFog, BuildFog, MetroFog or CCloud). As this phase carries the main data, this will affect the power consumption and the placement decision. Therefore, it is treated as the main component of the model. Note that processing the task locally in the mobile unit that has requested this service is not an option for the assumed scenarios. In the last two phases, the extracted knowledge resulting from the processed data is sent back from the processing node to the OLT, and so to the source mobile unit that requested the service. As we assume that the extracted knowledge has a small volume compared to the main data, the last two phases are not considered in the optimisation model.

In order to highlight the effect of the new integration between fog processing and optical wireless communication, the task assignment optimisation model has been evaluated here using an ideal scenario with eight mobile units located in the room. Each mobile unit is connected to a single light unit (access point) through one of the wavelengths (RYGB). Note that the sum of the OWC link capacities is restricted by the maximum capacity of the ONU connected to the access point, which is equal to 10 Gbps. This ideal scenario is compared at the end with the two scenarios resulting from the resource allocation model in Section 3.

We have evaluated the power consumption in the above-mentioned scenarios for an architecture composed of different fog servers in each layer. Based on the servers considered, the processing energy of the CCloud server is 82% more efficient than that of a mobile processor, followed by MetroFog, CampFog, BuildFog and then RoomFog server, which has just 32% of the processing energy efficiency of a mobile processor. On the other hand, because of the location of the CCloud and other fogs, the networking energy when traffic is sent to the RoomFog is 98% more efficient than with the CCloud. This is because the RoomFog server is only one hop away from the OLT controller, from which the tasks are offloaded. This networking energy increases by 1-3 mW/Mbps when the tasks are offloaded to the MobFog, based on the wavelength assigned to each mobile unit.

Table 3 summarises the capacities and energy efficiency values of the processing nodes and the route leading to each, while Table 4 provides the maximum capacity and power consumption values for each network device. Note that there is a single server in each of the five cloud and fog processing locations in Table 3. Therefore, the values of the networking efficiencies for different routes are defined there for each processing location.

The MILP model was evaluated to show the effects of the proposed architecture on the power consumption. We also studied the effects of different optical wireless wavelengths on the processing utilisation of the mobile units assigned. In this evaluation, 50 tasks were considered, with workload processing demands ranging between 100 and 1500 MIPS. In Equation (28), we introduce the relation between the workload demand and the data rate demand for each requested task as a ratio, termed the 'data rate ratio' (DRR). Different DRR values were defined in the model, varying between 0.002 and 0.6 to consider different scenarios with low and high data rates.

$$Data\ rate\ demand = DRR\ \times workload\ demand \tag{28}$$

The results are first presented for the ideal scenario, in which each mobile unit is connected to a single light unit with full, 10-Gbp data rate capacity through one of the wavelengths (RYGB). Then, results are given for scenarios 1 and 2, in which the eight users are assigned different wavelengths with varied data rates, based on their location. The different data rate values for each user in both of these scenarios are extracted from those given in Figure 5.

The results in Figure 7 show the processing power consumption, the networking power consumption and the overall power consumption versus the processing workload per demand for different DRR values.

**Table 3.** Processing nodes capacities and efficiencies

| Processing node (server) | Capacity (MIPS) | Efficiency (W/MIPS) | Route efficiency (W/Mbps) |
|---|---|---|---|
| Central Cloud (Intel® Xeon® E5-2680) | 144000 [98] | 0.000796 | 0.128 |
| MetroFog (Intel X5675) | 73440 [99] | 0.00129 | 0.0713 |
| CampFog (Intel Core2-Q9400) | 35160 [100] | 0.0027 | 0.0475 |
| BuildFog (Intel Xeon E5-2420) | 34200 [101] | 0.0028 | 0.0238 |
| RoomFog (Intel Core i7-6500U) | 6200 [102] | 0.003 | 0.0015 |
| Mobile unit (HTC One X) | 1500 [103] | 0.004 | 0.00222 (Red) |
| | | | 0.00195 (Yellow) |
| | | | 0.00177 (Green) |
| | | | 0.00177 (Blue) |

**Table 4.** Network devices capacities and power consumption

| Network device | Model | Power consumption | Capacity (Gbps) |
|---|---|---|---|
| OLT | Tellabs1134 [104] | 400 W | 320 |
| ONU | FTE7502 10G [105] | 15 W | 10 |
| Central cloud switch | cisco 6509 [106] | 3.8 kW | 320 |
| Central cloud router | Juniper MX-960 [106] | 5.1 kW | 660 |
| Core router | Cisco CRS-1 16-slots [107] | 13.2 kW | 1200 |
| Transponder (in core node) | ONS15454 [108] | 50 W | 10 |
| Optical switch | Cisco SG220 [109] | 63.2 W | 100 |
| Edge router | Cisco 12816 [110] | 4.2 kW | 200 |
| Aggregation switch | Cisco 6880 [111] | 3.8 kW | 160 |
| Ethernet switch | Cisco 6880 [111] | 3.8 kW | 160 |

The processing power consumption shown in Figure 7(a) indicates that at low DRR, the data rate is minimal and the networking power consumption becomes negligible. Hence, the location with the best processing energy efficiency is selected (the central cloud in DRR=0.002 and MetroFog in DRR=0.02). On the other hand, processing power consumption in DRR=0.04 and 0.06 starts linearly until certain workload demand (800 MIPS), as in Figure 7(a). After this point, a clear increase in the power consumption occurs when the most efficient locations, in total power consumption, start to become exhausted and the workload is consequently offloaded to other processing locations. It is also worth mentioning that in the case of DRR=0.04, the RoomFog becomes more efficient in processing the workload, followed by BuildFog and then MetroFog. At DRR=0.06, however, the MobFog performs better on energy efficiency than MetroFog due to its efficient networking energy. At higher DRRs, the processing power consumption of the MobFog increases with low and high increase (between 100 and 800 MIPS with DRR=0.2, 0.4 and 0.6). The increase varies according to the different workload demands assigned. For instance, with 300 MIPS, each mobile node can process five tasks with a total possible assignment of 1500 MIPS assigned to the mobile unit. This causes an increase in the processing power consumption, as the mobile units are the least efficient processors.

On the other hand, with 400 MIPS demand, three tasks assigned to each mobile unit have a total of 1200 MIPS assignment, which causes a reduced processing power consumption compared to the 300 MIPS case. These variable assignments are due to the limited capacity of the mobile unit and the single assignment constraint. Also, at the highest two DRR values, 0.4 and 0.6, the BuildFog and CampFog communication links become the bottleneck. Thus, processing is placed further out in the MetroFog, which reduces processing power consumption.

Figure 7(b), reveals a higher impact of the networking power consumption with increase in the workload demands at high DRR values. This affects the total power consumption as seen in Figure 7(c), which becomes highly related to the networking power consumption. Figure 7(b) shows that as DRR increases, ie as the communication data rate increases, the networking power consumption increases. At low DRR (0.002 and 0.02), the networking power consumption values are comparable and increase linearly with a very low power consumption. This is because the networking power consumption for such low rates becomes negligible with a minimum effect on the assignment decision. Consequently, the workload demands in these two cases are placed in one location (the location with the best processing efficiency).

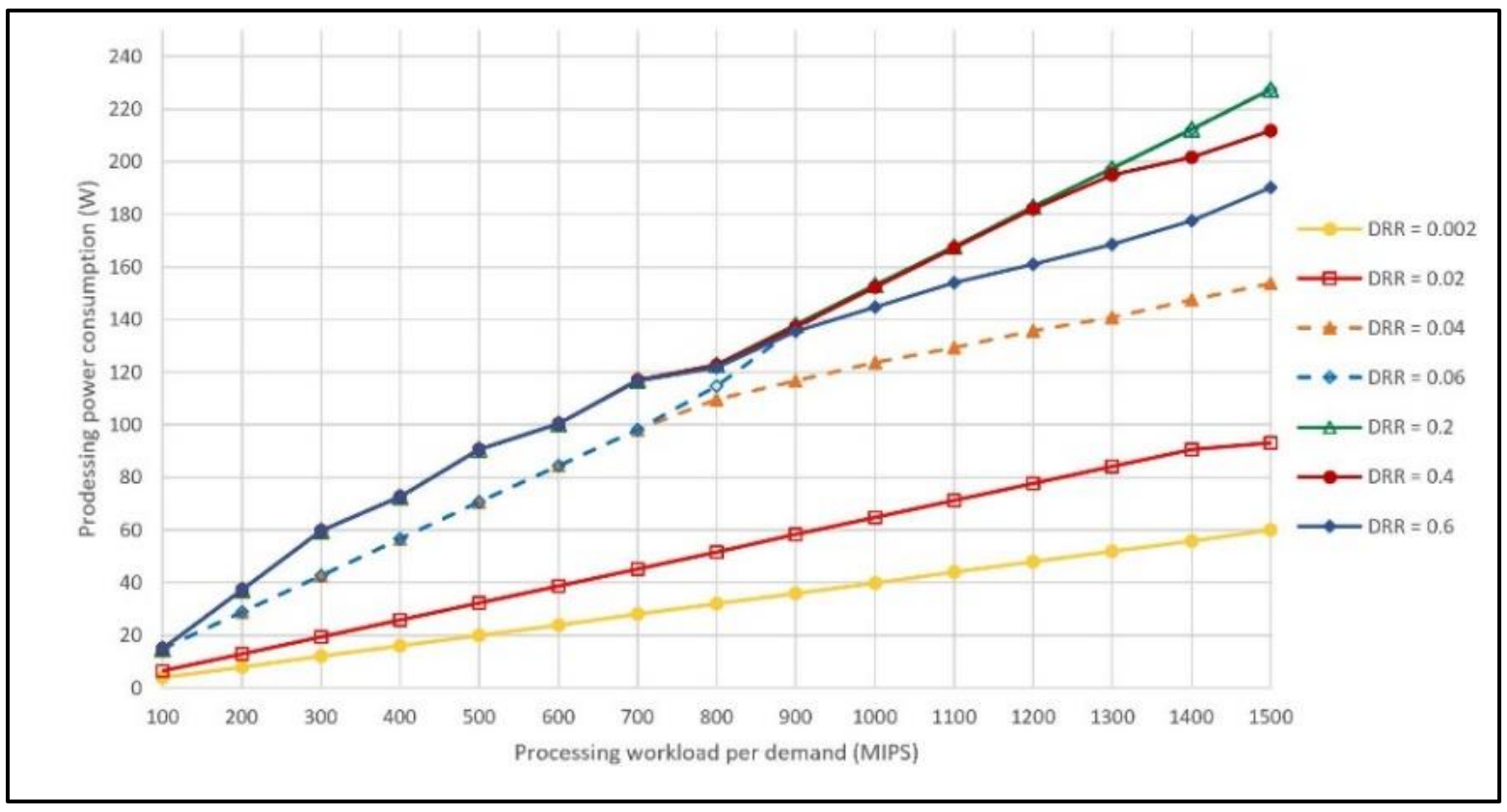

(a) Processing power consumption

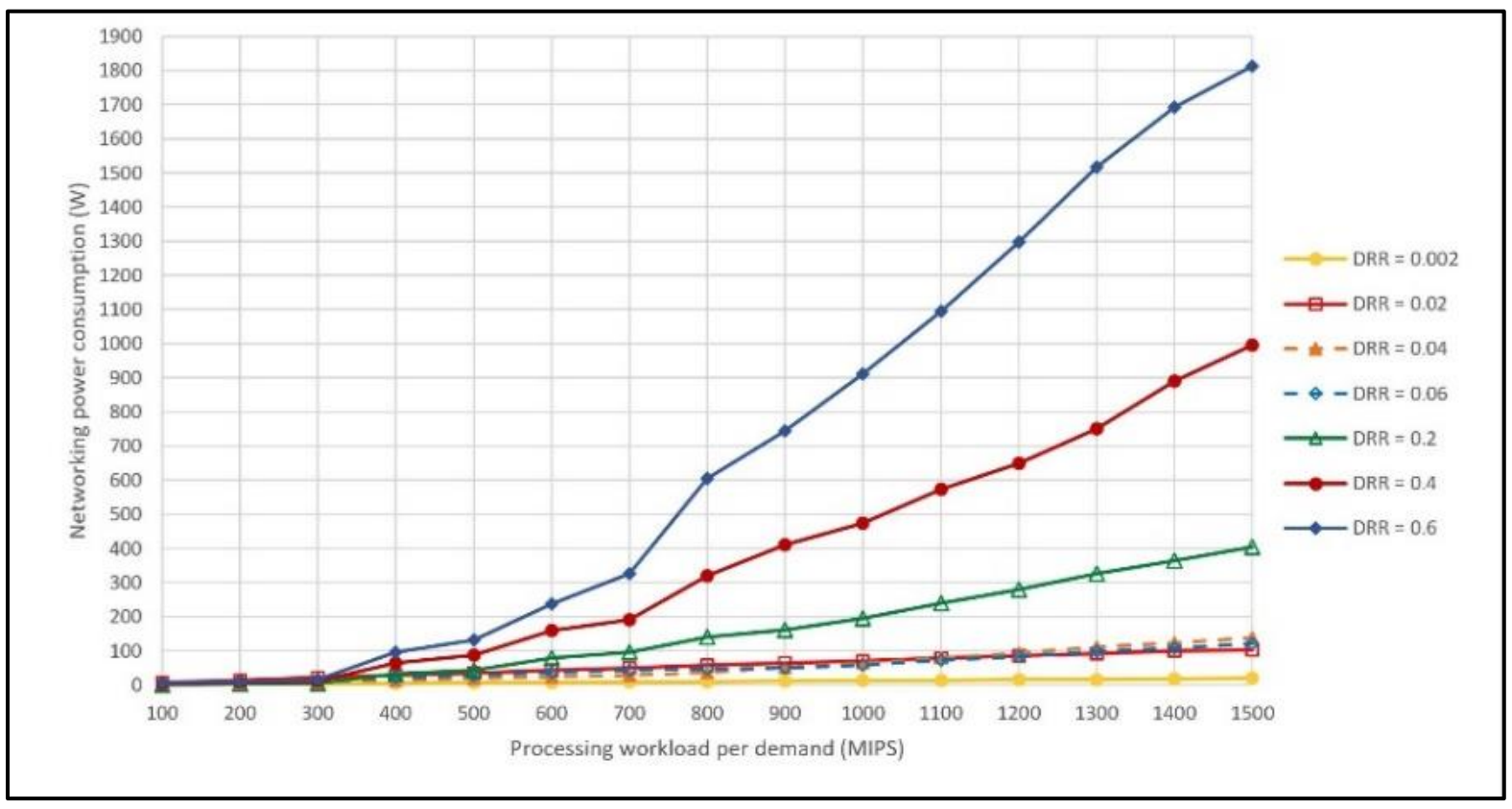

(b) Networking power consumption

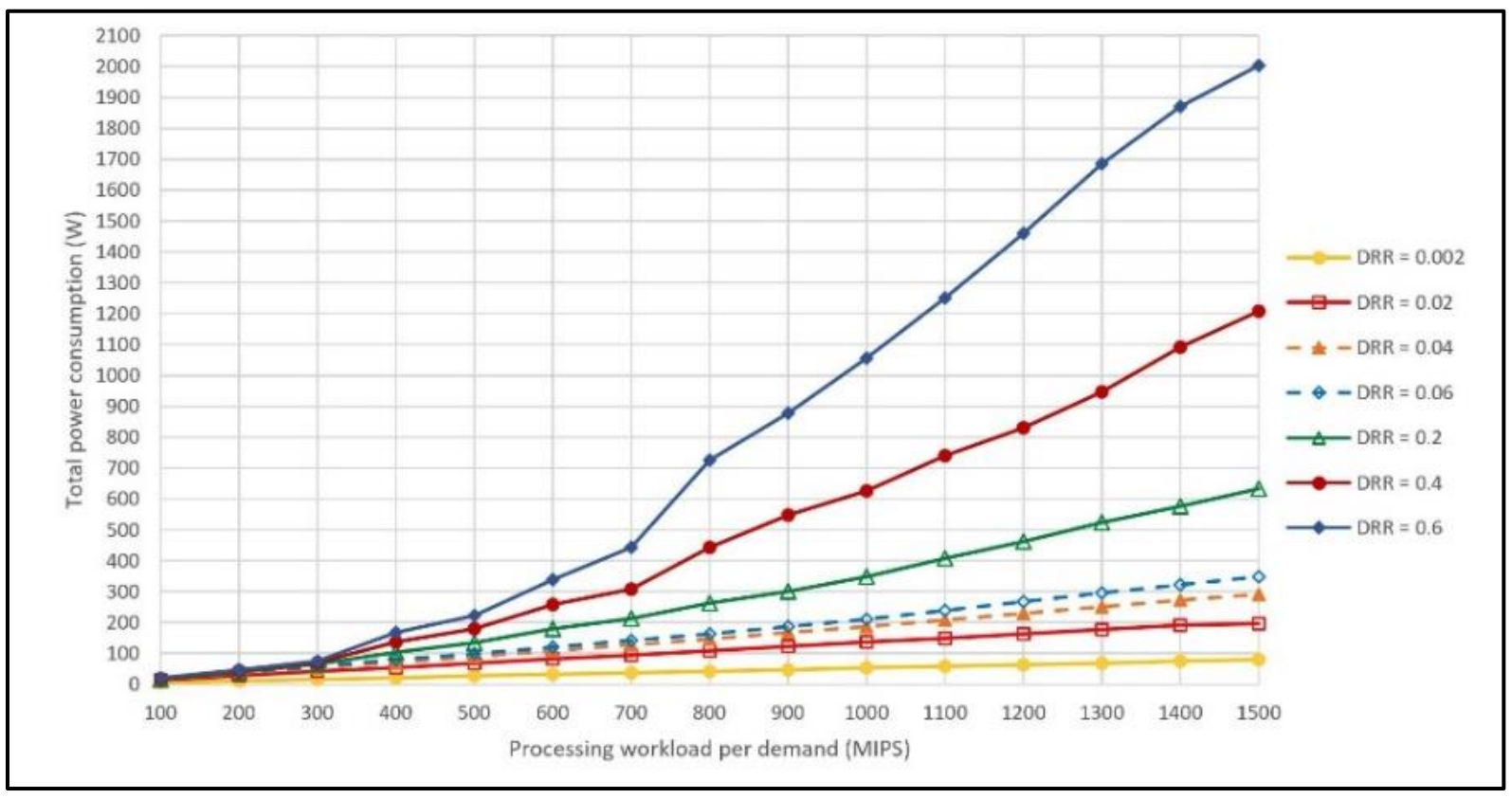

(c) Total power consumption

**Figure 7.** Processing, networking, and total power consumption vs. processing workload per demand, for different DRR values.

At DRR= 0.04 and 0.06, the placement decision remains affected by the processing power consumption. This results in a small increase at DRR=0.04 compared to that at 0.06 because the workload is offloaded in a further location (MetroFog) at 0.04 while at 0.06 it is offloaded to a more networking-efficient location (CampFog). A meaningful increase is shown at higher data rates, with DRR=0.2, 0.4 and 0.6. This power increases more significantly after a certain point (750 MIPS), when the mobile unit processor cannot process more than one task due to its limited capacity (1500 MIPS). In this case, more demands are offloaded to the CampFog, which increases the networking power consumption.

It is notable that at DRR=0.2, the power consumption has a slow increase as the tasks are assigned to the most efficient destination in terms of networking power consumption. In contrast, the power consumption increases more at DRR=0.4 and 0.6. The first reason for this is the limitation of the network links of BuildFog and CampFog, which cannot support very high traffic (in excess of 10 Gbps). Consequently, more tasks are offloaded to less networking energy efficient locations. The second reason is the limited processing capacity of the mobile, which results in the offloading of more demands further out.

Figure 8 shows the overall workload assignment for each processing location in relation to the networking power consumption for the case of DRR=0.6. For the assigned workload, the figure shows where demands are placed for each given data rate and how this affects the networking power consumption of this placement. Starting from low data rate demands, demands are assigned to the most efficient location in term of total power consumption (RoomFog followed by MobFog, BuildFog, CampFog and then MetroFog). Note that CCloud is not assigned any workload to satisfy the demand caused by any closed fog locations. Also, the networking power consumption for low data rate demands is very low due to the demands placed at RoomFog and MobFog. This power consumption is dominated by other fogs which are less efficient networking locations; thus, the networking power consumption becomes significant. More details about the networking power consumption splits for each DRR are shown in Figure 9.

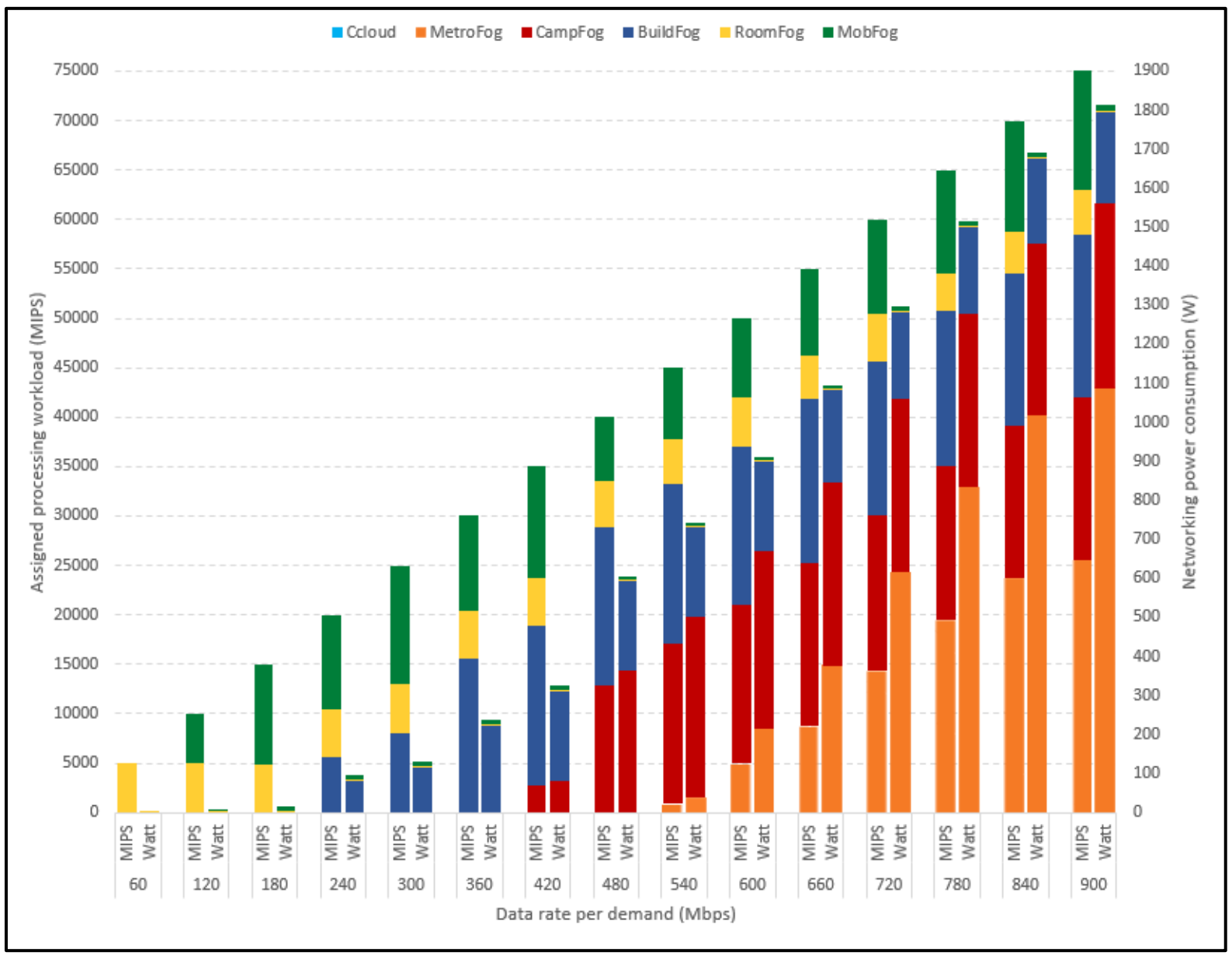

**Figure 8.** Assigned processing workload and networking power consumption vs. data rate per demand, for each cloud and fog node (DRR=0.6).

As mentioned (above), this ideal scenario considers a constant data rate wavelength assigned to each mobile unit. For scenarios 1 and 2, the data rate is limited based on the user location and number of users accessing the same access point, as explained (in the previous section). Figure 10 shows the processing, networking and total power consumption for these two scenarios (1 and 2). Scenarios 1 and 2 show comparable values for all power consumptions. This indicates that the demands assignment and therefore the consumed power are not affected by the different assigned channel data rates as long as these channels satisfy the networking demand and the mobile units satisfy the workload demand.

Figure 11 shows the mobile units processing utilisation based on the assigned wavelength for the three scenarios (ideal, 1 and 2). In scenario 1, each user is assigned a channel with data rate capacity ranging from 3.1 Gbps to 4.5 Gbps, while users are assigned channels ranging between 3.5 Gbps and 6.7 Gbps in scenario 2.

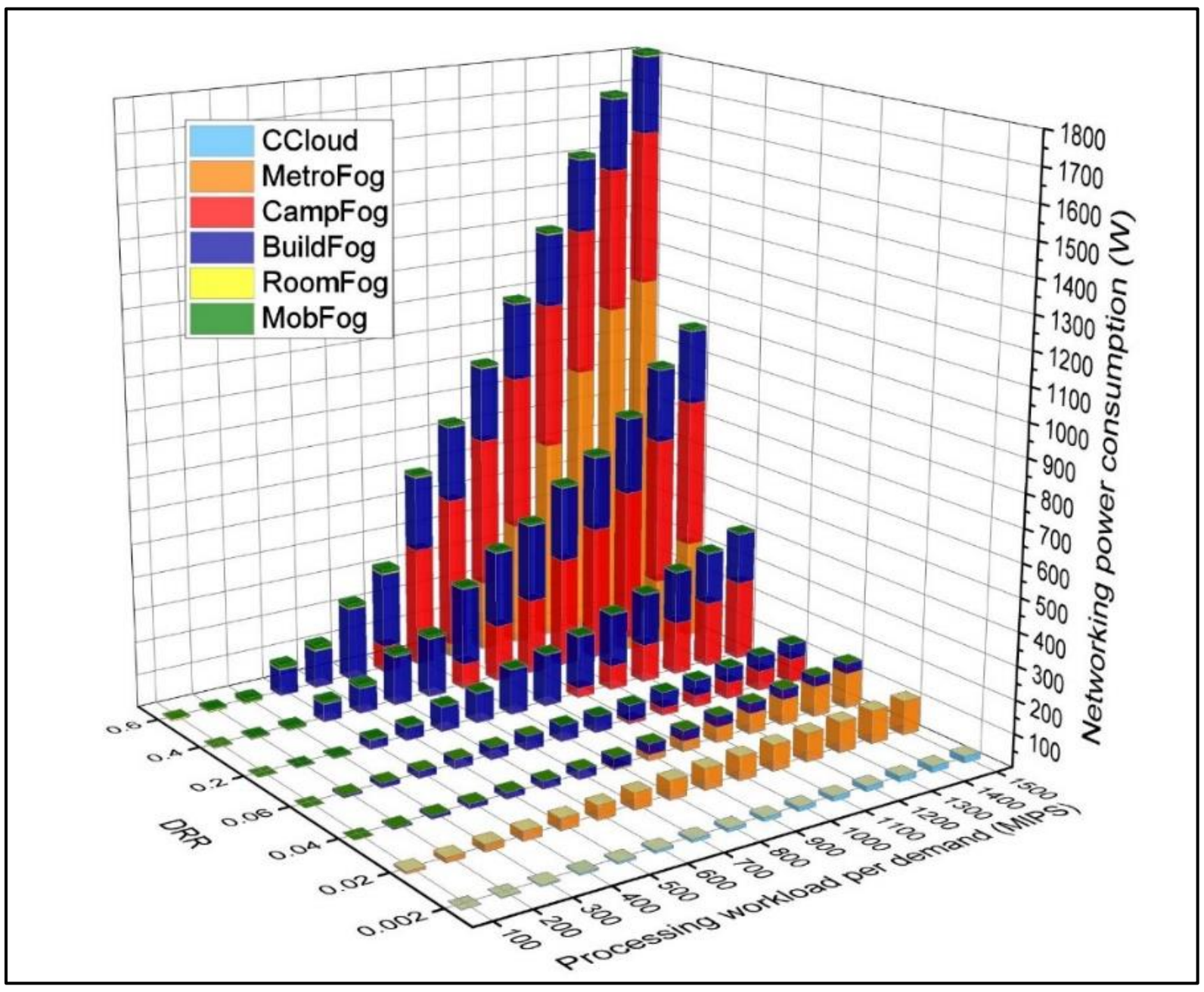

**Figure 9.** Networking power consumption vs. workload per demand and DRR value, for each cloud and fog node.

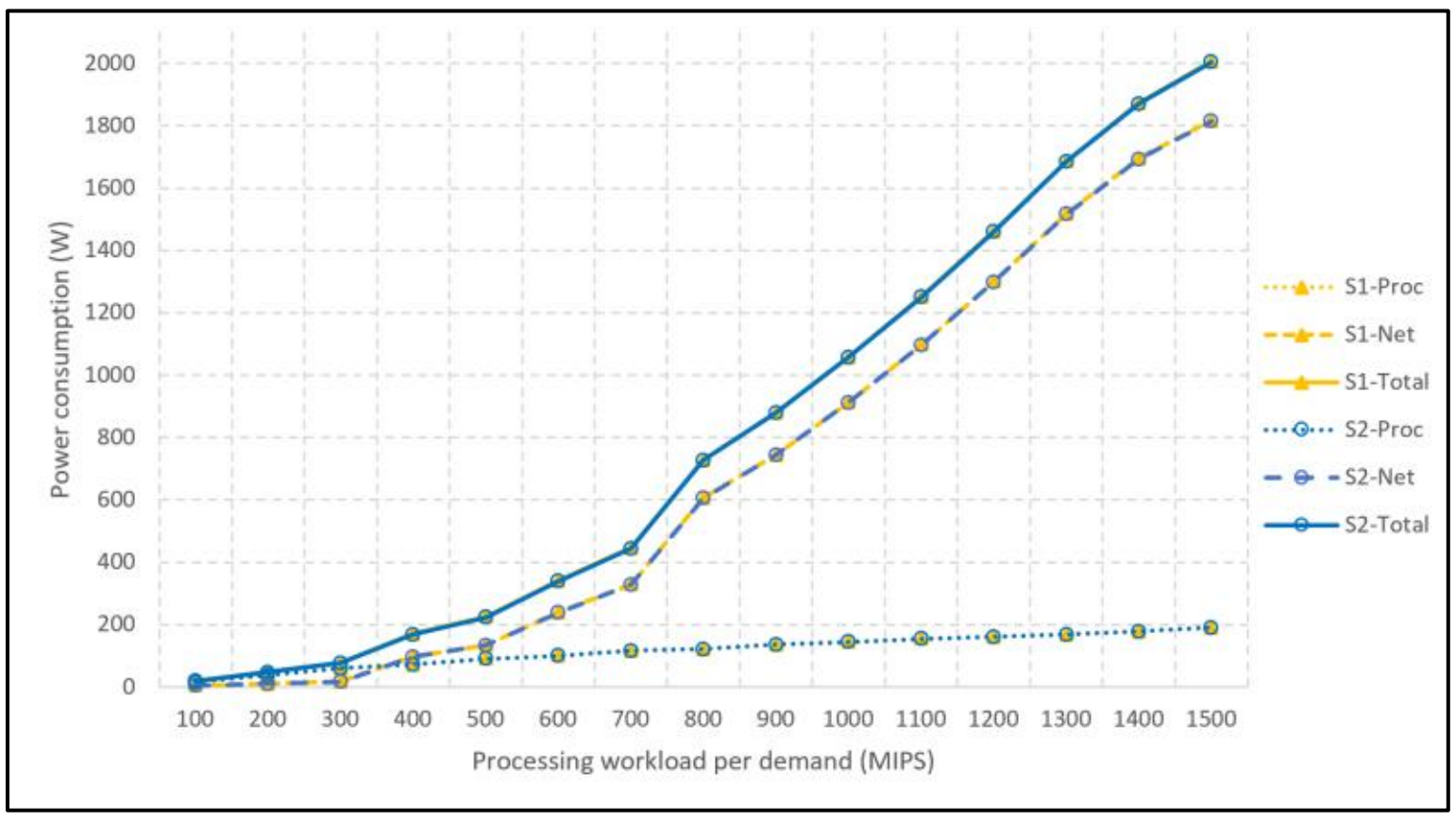

**Figure 10.** Processing power consumption, networking power consumption, and total power consumption, for scenario 1 (S1) and scenario 2 (S2) (DRR=0.6).

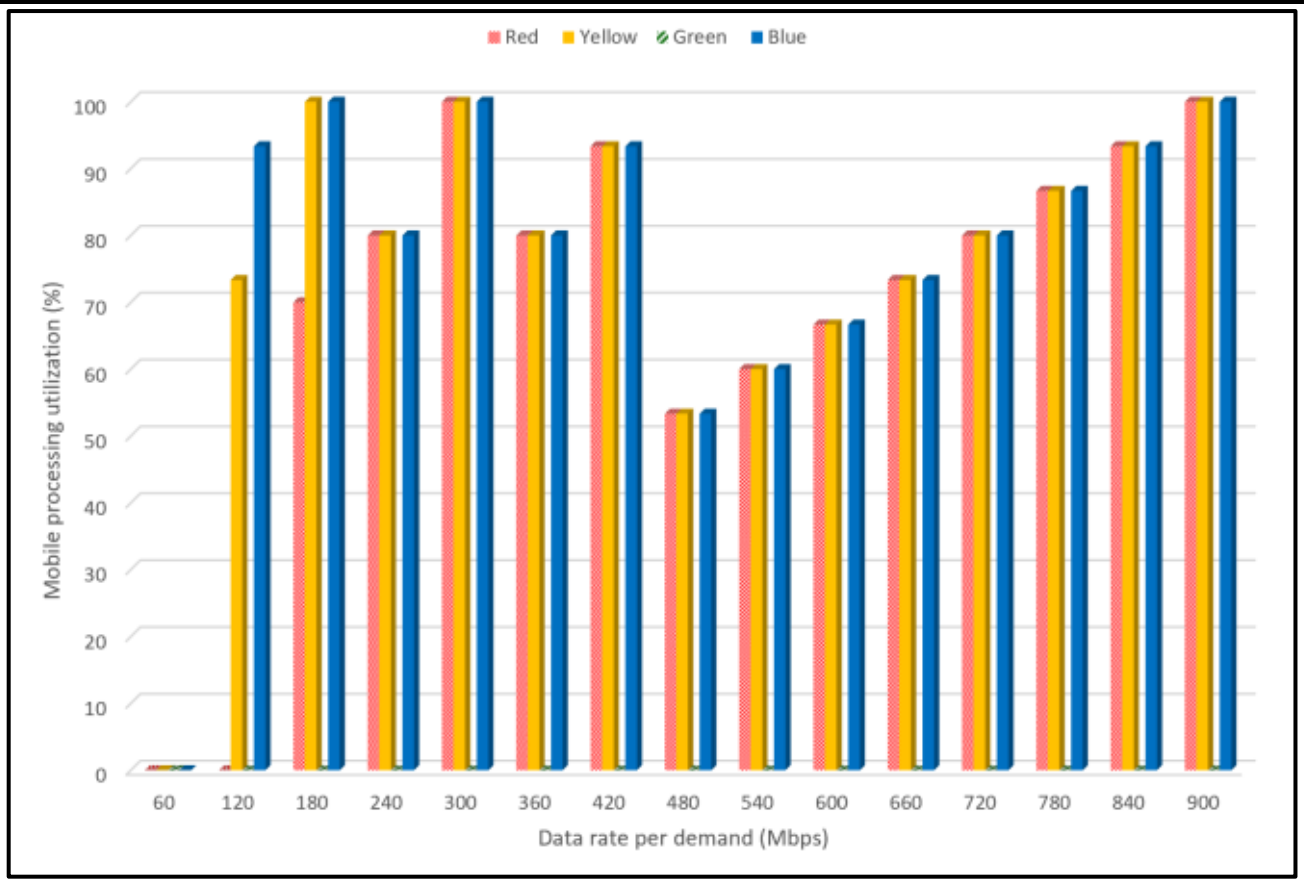

(a) Mobile processing utilisation in scenario 1

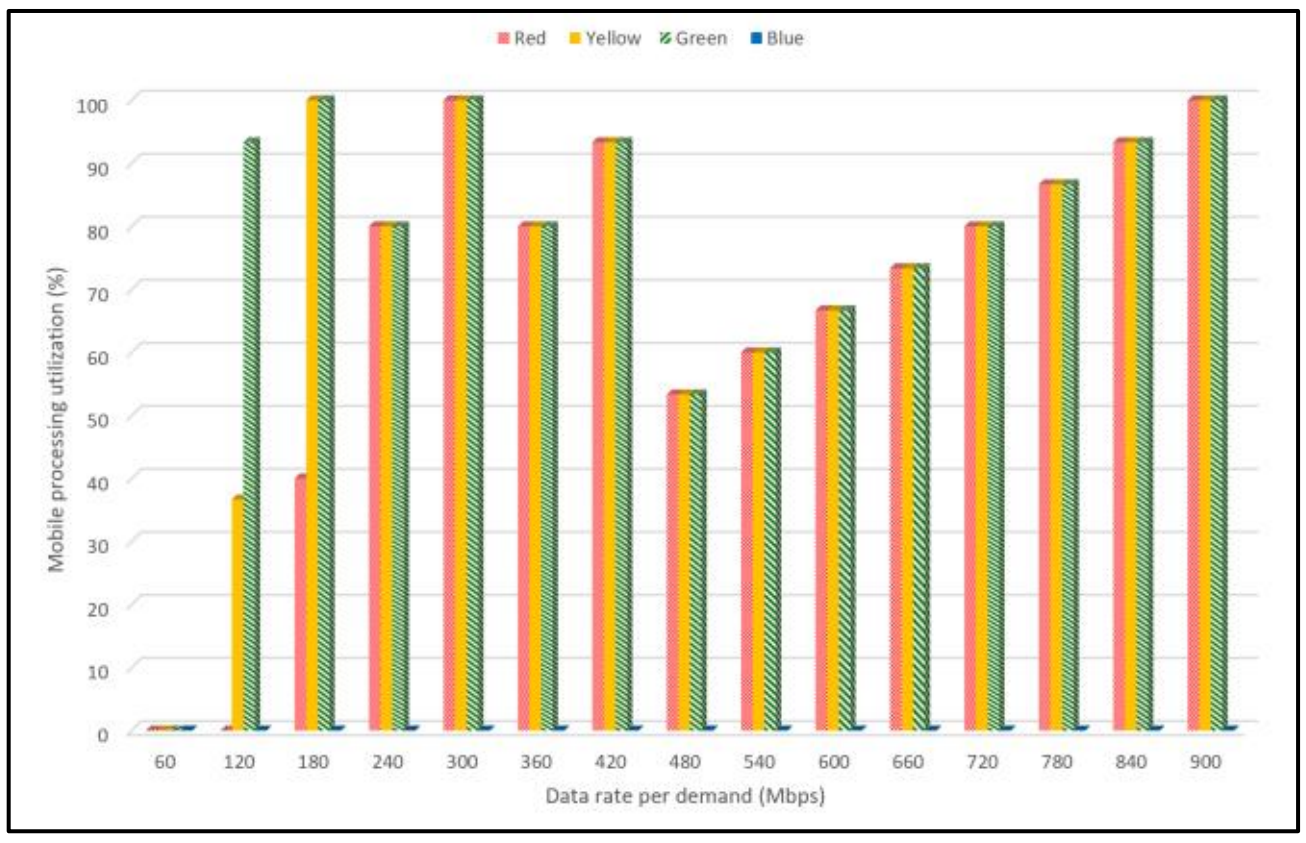

(b) Mobile processing utilisation in scenario 2

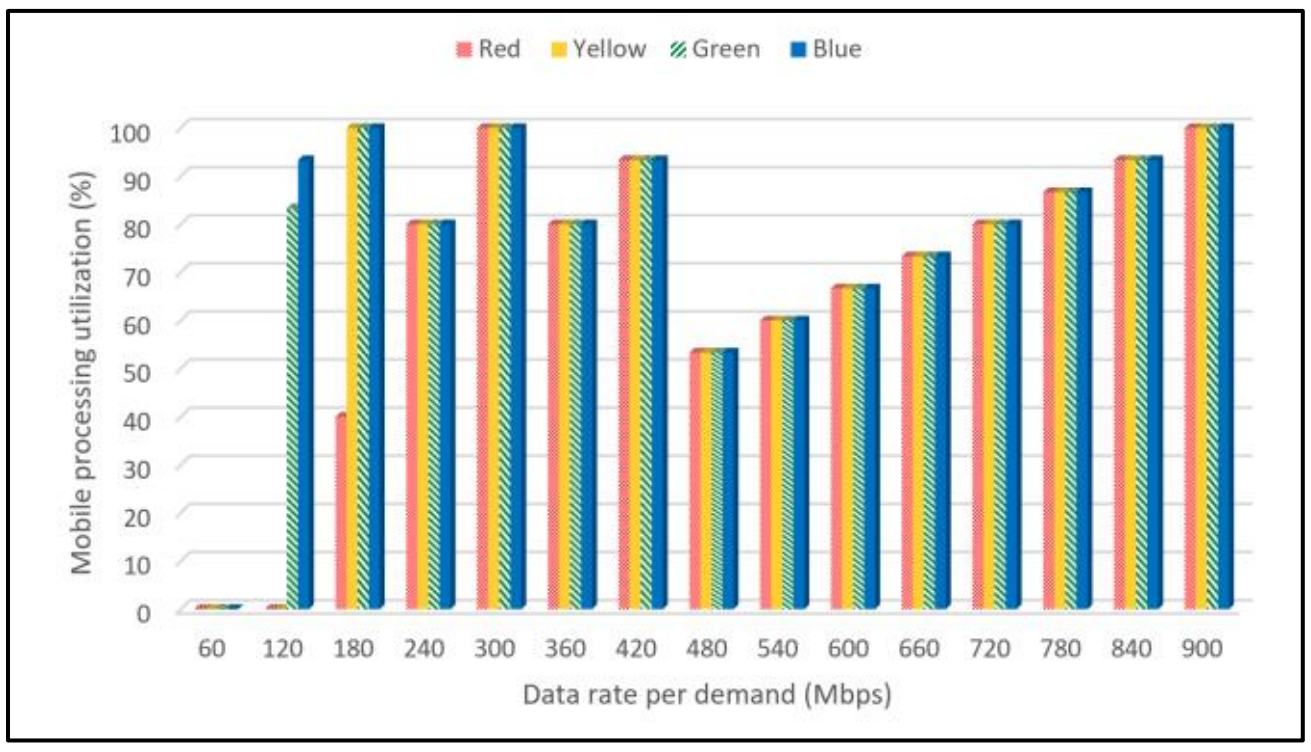

(c) Mobile processing utilisation in the ideal scenario

**Figure 11.** Mobile processing utilisation per assigned wavelength for scenario 1 (S1), scenario 2 (S2) and the ideal scenario (DRR=0.6).

Figure 11(a) shows that the mobile unit utilisation is equal to zero at the lowest data rate demands as all the workload demands are assigned to the RoomFog. At 120 Mbps, the processors of mobile units assigned to blue wavelength are occupied first as these have the lowest wavelength energy (see Table 2), followed by the mobile units assigned to the yellow wavelength. At 180 Mbps, mobile units assigned blue and yellow wavelengths achieved their full utilisation, followed by mobile units assigned to the red wavelength which has the highest energy usage (here, in scenario 1, there are no mobile units assigned to the green wavelength). At 240 Mbps, all mobile units achieved only 80% utilisation due to bin packing and single assignment considerations. For instance, based on the DRR value (0.6), the workload demand at 240 Mbps is equal to 400 MIPS for all 50 tasks.

With a mobile processor capacity equal to 1500 MIPS, each mobile unit can serve three tasks with a maximum assignment equal to 1200 MIPS per unit. However, full utilisation is achieved at 300 Mbps as three tasks are assigned to each mobile unit with a total assignment that is equal to the mobile processing capacity. At 360 Mbps, the utilisation decreased to 80%, due to the same bin packing and single assignment considerations. This utilisation increases to 93% at 420 Mbps as the demand workload is equal to 700 MIPS so each mobile unit is assigned two tasks with a total assignment equal to 1400 per unit. A significant utilisation drop occurs afterwards, at 480 Mbps, as the workload per demand increases to 800 MIPS. This is due to the limited processing capability of the mobile unit, which cannot serve more than one task, causing the mobile processor to achieve only 53% utilisation. The constant increase in the workload demand causes a constant increase in the mobile unit utilisation, which ranges between 60% (at 540 Mbps), and 100% (at the highest considered workload demand, 1500 MIPS). Similar observations apply to scenario 2 (where no mobile units are assigned the blue wavelength) and the ideal scenario, as shown in Figures 11 (b) and (c), respectively.

# 5. Conclusions and Future Work

In this paper, an OWC system was used to support multiple users. A wavelength division multiple access (WDMA) scheme was used to support multiple users served simultaneously by the OWC system. A MILP model was developed and used to optimise resource allocation and was shown to increase the system throughput and allow multiple access. Thereafter, a cloud/fog-integrated architecture was built to create a connection with potential mobile nodes and to provide processing services for these mobile nodes. The mobile OW nodes were also clustered as fog mini servers and were thus also able to provide processing services to each other. A second MILP model was proposed to optimise the processing placement by minimising the total power consumption. Future areas of work can include (i) consideration of the uplink as the current work only considered the downlink; (ii) consideration of additional wavelengths for multi-user support enabled through infrared and WDM for example. In the current system, multiple access using WDM, was limited by the four wavelengths available in VLC; (iii) consideration of additional multiple access dimensions beyond wavelengths. These can include TDM, OFDMA, orthogonal coding, spatial diversity and beam steering which can reduce interference between users; (iv) consideration and minimisation of latency in addition to the minimisation of power consumption when allocating resources and placing processing jobs; (v) development of heuristics building on the MILPs insights to enable real time resource allocation and task placement (vi) additionally, as this is a new integration between optical wireless and fog computing, more work is needed to address virtualisation, the software matching problem (where only a subset of the processing nodes have the relevant software for task placement), handover and quality of provided services. Also, more opportunistic scenarios need to be evaluated with other forms of fog computing.

# Additional Information

**Data Accessibility**

All data are provided in full in the results section of this paper.

**Authors' Contributions**

JMHE Conceived the concept and suggested the multiple access MILP optimisation and processing optimisation and improved the models and results. OZA and MTA developed the optical wireless channel models and optical wireless channel modelling tool, obtained the optical power and interference results. SOMS developed the wavelength and access point resource optimisation MILP and results. AAA developed the network architecture and processing placement optimisation MILP and results to minimise power consumption. SHM and TEHE helped with the development of the two MILP models and verified them.

**Funding Statement**

This work was supported by the Engineering and Physical Sciences Research Council (ESPRC), INTERNET (EP/H040536/1), STAR (EP/K016873/1) and TOWS (EP/S016570/1) projects. The authors extend their appreciation to the deanship of Scientific Research under the International Scientific Partnership Program ISPP at King Saud University, Kingdom of Saudi Arabia for funding this research work through ISPP#0093.

**Acknowledgments**

OZA would like to thank Umm Al Qura university in the Kingdom of Saudi Arabia for funding his PhD scholarship, AAA would like to thank the Imam Abdulrahman Bin Faisal University in the Kingdom of Saudi Arabia for funding her PhD scholarship, SOMS would like to thank the University of Leeds and the Higher Education Ministry in Sudan for funding her PhD scholarship. SHM would like to thank EPSRC for providing her Doctoral Training Award scholarship.